\documentstyle[12pt]{article}

\newcommand{\beeq}{\begin{equation}}
\newcommand{\eneq}{\end{equation}}
\newcommand{\beeqar}{\begin{eqnarray}}
\newcommand{\eneqar}{\end{eqnarray}}
\begin{document}
\vskip 0.5in

\title{ An Elementary (Pseudo)Scalar at all scales.}
\begin{center}
An Elementary (Pseudo)Scalar at all scales.\\
\vskip 0.5in
N.D. Hari Dass\dag\\ 
Institute of Mathematical Sciences, C.I.T Campus, Chennai 600 113, India.\\
V. Soni\ddag\\
National Physical Laboratory, New Delhi 119912, India.
\end{center}
\begin{abstract}
We consider an extended linear $\sigma $ model in which the fermions are
quarks and are coupled to gluons. Equivalently, this is QCD extended by
coupling the quarks to a colour singlet chiral multiplet of ($ \sigma,
\vec \pi $) fields. This theory has a phase governed by a UV fixed point
where all couplings are AF (asymptotically free). This implies that the
scalars are elementary at high energies (UV) and, as they are colour singlets,
they are not confined at low energy (IR ). Thus , the scalar
particles are elementary at all scales.
\end{abstract}

\section{Introduction}
\noindent
The question of elementarity of a particle has been 
an abiding theme in Physics.
From the atom to the nucleus, to the nucleon, all 
have been treated as
elementary till proved otherwise. The higher the 
energies we explore, and
this is a process that goes monotonically with time, the greater the
resolution. It is, then, not surprising that elementarity has been a strongly time dependent attribute.
However, we do have some long term 'elementary' survivors like the electron
and the
quark, to name a few. The electron has continued to be elementary from
the time it was discovered to this day whereas the quark is a recent arrival.
There is a rather important difference , the electron is manifest in a free
state but the quark is not - it is  confined.

From  a field theory point of view this difference between the quark and the
electron is perhaps even more interesting.
We could consider the electron 
entirely within the framework of
QED, which is not an asymptotically free theory 
whereas the quarks
can be described by QCD which is an asymptotically 
free( AF ) theory. In that case 
the electron couples to the photon more and more strongly as the energy goes
up  and so we cannot predict its fate ( or its elementarity) at some very
high energy when the
coupling diverges at the so called Landau Singularity(LS).
In QCD, the quark coupling to the gluons , however, decreases as we go to
higher and higher energies and so it starts behaving like a free particle or
parton and is in no danger of losing its elementarity as the energy goes
higher.

 So it seems that particles that are strictly governed 
 by a non abelian gauge
 AF theory can be elementary in the sense that they cannot resolve any further into
 constituents as the energy goes up. This is a new way to look at
 elementarity.

  This, however, comes at a cost. Particles of a non 
  abelian
 gauge theory, like quarks and gluons, turn out to 
 be confined and so cannot
 be seen in a free state. This means that these particles are absent from the
 spectrum of the theory. It is not meaningful, then, to talk about
 their elementarity at these scales.

 This leaves us in a bit of a bind. Particles which 
 belong to an AF theory like QCD
 are clearly elementary at asymptotic energies but 
 are confined at low
 energy. At the other end particles that are seen 
 only at low energies are
 generally composite. Can we find a particle that 
 eludes this catch i.e one which is
 elementary at all scales ?

 In what follows, this problem is eluded by considering an extension of
 QCD by the addition of a colour singlet chiral 
 multiplet,
 ($ \sigma, \vec \pi $),
 that couples to quarks via a yukawa
 coupling. This chiral multiplet should not be confused
 with the dynamically generated multiplet that consists
 of the Goldstone boson, the pion. This coupling, as it turns out, can be AF 
\cite{primo,secundo}. The AF
 nature of the yukawa coupling is possible only when the
 quarks are also coupled to an AF non
 abelian gauge theory. Since it is the AF nature of the yukawa coupling that
 makes the scalar  elementary, the  elementarity of the scalar can  only be
 gained by riding piggy back on a nonabelian gauge theory.

Further, since the scalar is not confined like the the quarks, it is
elementary
and visible at all energies! This is indeed a new situation in
Particle Physics. The electron in QED has a problem 
at very high energy, when the
coupling gets large and has a Landau Singularity
and we have no way
to handle the physics \footnote { Of course, this is evaded when QED is
embedded in a larger AF theory.}. The quark  or gluons, on the other hand, are confined
at low energy and so the question of their elementarity at low energy is not
very meaningful.

The issue of elementarity vs compositeness in local 
field theory has a long and interesting history. An
important first step was the construction of a local
field operator for composite particles \cite{haag,nishi,zimmer}. A central idea in the field theoretic context
was that compositeness is signalled by the vanishing
of the wavefunction renormalisation. While this 
remained a conjecture for sometime which was supported
in some specific models \cite{jouvet,howard,vaughn},
Fried and Jin on the one hand \cite{jin}, and 
independently Divakaran \cite{ppd} on the other, were
able to prove this on general grounds. Divakaran was 
also able to demonstrate the so called "indifference
hypothesis" of Feynman according to which the 
Lagrangian can be written as a function of {\it any}
one of all the irreducible sets of fields. In
particular, sets including the composite {\it fields}
can also be used. This immediately indicates that
characterisations according to which elementary 
particles are those whose fields appear in the
Lagrangian and composite the ones whose fields do not 
appear so \cite{vaughn} are rather naive. This fact is
explicitly borne out in the works of Hasenfratz et al
\cite{hasen} and Zinn-Justin \cite{zinn}.

Coming back to the question of wavefunction
renormalisation,
if $Z_\phi $ vanishes at some scale, it implies that 
the particle loses its kinetic term and 
that it is composite
from  this scale onwards. However, often it is not 
possible to compute $Z_\phi $
except in perturbation theory - making it difficult 
to extend the
calculation all the way to the region where 
$Z_\phi = 0 $. The criterion is therefore
more useful
as a rough guide to the compositeness scale. 
As we point out, $Z_\phi = 0 $
often occurs simultaneously with the coupling diverging at a LS.
In the case of confinement, we expect $Z_\phi = 0 $ at 
the confinement scale, which is inherently 
non perturbative and at present not calcuable.

Even within the framework of 
S-matrix theory it is interesting to raise the
distinction between elementarity and compositeness. At
first sight it would appear that since both of them 
manifest as poles no clear distinction would be 
possible. Within the Lee model \cite{lee} Vaughn et
al \cite{vaughn} indeed found a distinction that would
be reflected in the properties of the S-matrix alone.
This was in the high energy behaviour of the scattering
phase shifts similar to the case in potential 
scattering as indicated by the Levinson \cite{levinson} theorem.
Rajasekaran \cite{graj} has proposed another way of 
distinguishing elementary and composite particles within S-matrix theory; according to him absence of a pole
in the K-matrix signals compositeness.

We have scenarios of dynamical symmetry breaking 
(see for example, Refs.{\cite{bardeen},\cite{miransky})}) where a sufficiently strong
effective four fermion interaction at the compositeness
scale can generate kinetic terms for a new particle mode below the
compositeness scale 
and also  spontaneously break an explicit
symmetry which will convert some particle modes to  goldstone bosons
( for example the QCD `pion').
In these models the low energy theory is a
renormalizable sigma model. Using perturbative 
renormalization  group (RG)
 it is found that as we approach the  scale of the
 LS at which
the yukawa coupling diverges, $Z_\phi  $ ( for the scalar
 /pseudoscalar ( PS )) simultaneously goes as an inverse power of the yukawa
coupling. As we approach this scale we may use the approximation of dropping
the kinetic term for the scalar. In this circumstance,
the four fermi interaction is recovered
using the field equation. This
scale may be loosely identified with the compositeness 
scale for the scalar
 (PS). Evidently, this procedure is far from rigorous, since the analysis
breaks down for large coupling long before the LS for the yukawa coupling is
reached. The above cosiderations are then only a pointer to compositeness and
not a proof.
\subsection { Elementarity and fixed points. }
\noindent
The question of fixed points is of fundamental
importance in understanding the behaviour of a field 
theory; in
particular of the scaling properties of a quantum field theory.

In four dimensions Yang Mills gauge theories are the 
only theories that have
a non trivial ultra violet (UV) fixed point 
\cite{gross}. 
The
presence of a UV free fixed point for this theory is equivalent to the
property of asymptotic freedom (AF) which, in turn, translates into the
scaling behaviour of scattering amplitudes.

On the other hand, the $ \lambda \phi^4$  field theory 
( nonperturbatively ) is
governed by {\it only} a gaussian IR fixed point and is
therefore trivial. Perturbatively
, as we know, the coupling diverges at LS.

The conventional wisdom on scalar field theories is that they are generally not
AF.  The introduction of fermions that
interact with the scalars, e.g. the linear $ \sigma $ model of Gell Mann
and Levy, makes
these theories more interesting but nevertheless it has been shown that they
cannot be asymptotically free \cite{gross}. This implies that
they must
be theories defined by a finite cut off.  However, this does not clear up
the issue of their triviality; in the sense that as the UV cut off is removed
it is not clear that such theories go to free field theories. It is
conceivable that these theories may be governed by a non gaussian  IR fixed
point.

Hasenfratz et al \cite{hasen} and Zinn-Justin \cite{zinn} consider a large N
yukawa field theory in which the fermions (quarks) are $ N$ component colour
fields that couple to scalar fields. The colour index 
is a free index as there
are no colour interactions. Effectively, the N components  perform  a counting
operation to permit a large N expansion.
They find that these theories are governed by an IR gaussian fixed point in
analogy to the  $ \lambda  \phi^4 $  theory.
They also find that the Nambu-Jona-Lasinio model which has the same
symmetry but is perturbatively non-renormalizable, is also governed by the
gaussian IR fixed point in the large N limit.

An interesting question then arises as to whether the introduction of non
 abelian
gauge fields that couple to the fermions can change the fixed point structure of
the theory.

In \cite{primo,secundo}
we considered a chirally invariant Yukawa-gauge theory 
in which a colour singlet, flavour chiral multiplet, 
was coupled to quarks in QCD, for the case $N_C =3 $ 
and $ N_F =6$. Another such theory in which
the degrees of freedom are quarks, gluons, Higgs bosons,
weak vector bosons and
leptons is the standard EW model (the GSW model)
\cite{schrempp,harada}.
We have shown for such theories that:
 i) if the initial data on the couplings falls in a certain region
in the parameter space of the couplings, namely , $\rho = g_y^2/g_3^2 
 < \rho_c $ where $ g_y$ and $g_3$ are the yukawa and QCD
 couplings respectively ($\rho_c $ is determined by  
 $N_C  $ and $ N_F $), and, ii) only on a
specific trajectory (the Invariant Line, IL) in the [R,$\rho$] parameter
space (where  $ R = \lambda/g_y^2 $, and $\lambda$ 
the scalar self coupling),
the theory can be asymptotically free. For couplings outside
of this specific region the theory is not asymptotically
free.

The AF branch above gives us a chiral invariant and 
completely asymptotically
free theory (that is AF for all couplings), which 
can be a candidate theory
for the strong interactions, similar to QCD. 
In \cite{secundo},
the difference between this theory and QCD is made 
explicit. Due to AF the
entire spectrum of this theory, including the colour 
singlet scalar, is parton like or elementary at 
high energy.

This result runs counter to the popular belief that a generic non abelian
guage theory with scalars may not be AF.
This  demonstrates that when we couple non abelian 
gauge fields, gluons, to a chiral theory of quarks 
and scalars the theory can have an
entirely new fixed point - an UV  fixed point, as 
in QCD. This is in
contrast to the case in the absence of the gauge 
fields, when the theory
is governed by the IR free fixed point alone. 

\section{ The theory}
\noindent
We begin with a spontaneously broken Yukawa gauge 
theory given by the Lagrangian below \cite{primo,secundo,long}:
\beeqar
{\cal L} = &- &\frac{1}{4} G^a_{\mu v} G^{a\mu v}
|_{colour} - \sum {\overline{\psi}} \left( D + g_y(\sigma + i\gamma_5
\vec \tau\cdot\vec \pi)\right) \psi - \frac{1}{2} (\partial_ \mu \sigma)^2 - \frac{1}{2}
(\partial_ \mu \vec \pi)^2\nonumber\\ 
&-& \frac{1}{2} \mu^2 (\sigma^2 + \vec \pi^2) - \frac{\lambda}{4}
(\sigma^2 + \vec \pi^2)^2 + \hbox{const} 
\eneqar
We consider here a particular chiral theory. The 
fermions are quarks.
The scalars 
($ \sigma, \vec\pi $) 
are a {\it colour singlet} chiral multiplet of
fields.
The quarks  belong to the
fundamental representation of the colour group
$ SU(N_C) $.  All quark generations
couple to the chiral multiplet identically with an 
$ SU_2(L) \times SU_2(R) $ symmetry.

We  find that when $ N_C $ and $ N_G $
are mutually constrained and so also the ratio of 
the gauge and Yukawa couplings, this theory can be 
completely asympototically free (AF) for all
couplings. It should however be noted that the
various couplings need not maintain fixed ratios 
with respect to each other.

We give below the 1 loop beta functions of this theory:
\beeq
\frac{\partial{ g_3(t)}^2}{\partial t} = - g_3^4/8\pi^2 P 
\eneq
where
\beeq
P = 11N_C/3 - 4 N_G/3. 
\eneq
Likewise
\beeq 
\frac{\partial g_y(t)^2}{\partial t}  = \frac{1}{8\pi^2} g_y^2
(4 N_G N_C  g_y^2 - 3 g_3^2(N_C^2 -1)/N_C ) 
\eneq
Note, that in our case $ N_F =  2N_G $. These equations
can be brought to the form
\beeq 
\frac{\partial \rho}{\partial g_3^2} =  - \frac{ \rho }{2 g_3^2 P } 
 L 
\eneq
with
\beeq
 L  =  8 N_C N_G \rho + 2P  - 6 (N_C^2 -1)/N_C 
\eneq
where $\rho$ is as defined before.
The zero of L occurs at $ \rho  = \rho_c $ where
$\rho_c$ is given by
\beeq
\rho_c  =  \frac{4( 2N_G -N_C)/3 -  6/N_C}{ 8 N_C N_G} 
\eneq
Note that the derivative of $ \rho $ above is positive 
when L becomes negative, i.e. for $\rho < \rho_c $.
In this case $\rho$ increases as $g_3^2$ increases.
Since $ g_3$ decreases with increasing momentum scale 
(since it is AF), this means that $\rho$ will also 
decrease with increasing momentum scale. The implication
then is that if $\rho < \rho_c $ then $ g_y^2$ is also 
AF and decreases faster than  $ g_3^2$ in the UV. It is
then clear that the the behaviour of $ g_y^2$ is now 
controlled not by the IR free fixed point \cite{hasen,
zinn} but by the UV free fixed point that controls the 
AF QCD coupling. This is possible only if 
$N_G > (N_C/2 + 9/(2 N_C))$.

On the other hand AF of $ g_3 $ requires P to be 
positive, i.e. $N_C > 4N_G/11$.
The constraint required to have both an AF $ g_3^2$ 
as well as an AF $ g_y^2$ is, for
large $ N_C $, $2 N_G >  N_C > 4N_G/11$.

Moving on to the scalar self coupling, the RG flow
equation is
\beeq
\frac{\partial \lambda(t)}{\partial t} = 2N_G (4 \lambda N_C  g_y^2 +
 +  12 \lambda^2/(2 N_G)  - 4 N_C g_y^4)/8\pi^2
\eneq
After a little algebra this can be brought to the
form
\beeq
\frac{\partial R}{\partial \rho} = \frac{4 N_G}{L} 
\left( 2R - 4 + R \frac{3}{2\rho} \right)
\eneq
with $R$ as defined before.
As observed in \cite{schrempp,harada,long},
for the constraints given above it has three quasi-fixed
points which are also the zero slope points in the 
above equation. These quasi-fixed points are given by:
$i) \rho = 0, R = 0;~~~~ii) \rho = \rho_c, R( \rho_c); 
~~~~iii)  \rho = \infty,
R = 2$.

The above equation is characterised by an invariant line
(IL)\cite{ross,Zim,KSZ} that goes
through all the above points. It divides the paramater
space into two regions (see \cite{schrempp,long} 
for details):

i) {\it The AF phase : $\rho < \rho_c$}. In this region the IL is the only
trajectory on which an AF theory in all couplings can be defined.
Wherever we are on the IL, we move to the origin as we go to the UV.
All other trajectories diverge. There is a reduction in coupling
constants on the IL\cite{Zim,KSZ}; $ \lambda $ is fixed
once $g_y$ is fixed. Around the
origin, it is easy to see that R is proportional to $ \rho  $ on the IL.
It is then clear that as R goes to zero in the UV , that  $ \lambda $
goes to 0 even faster than $ g_y^2 $ in the UV.

This completes the demonstration that the coupling, $ \lambda $, is also AF
on  one trajectory ( IL ) and governed by a UV free 
fixed point. Since all
couplings are AF, the entire spetrum, including the 
scalars, is elementary at high energy.

ii) {\it The non AF phase: $\rho > \rho_c$}. In this region the theory
is not AF in the yukawa and scalar self couplings. The fixed point structure
is rather complicated and the scalars are not 
elementary. This phase will be considered separately.

\subsection{Remarks}
\noindent

 1. Till now we have considered the RNG evolution equations in the absence of
the electroweak  gauge coupling. The effect of the weak
gauge coupling is to increase the value of $ \rho_c $,
permitting us to find a completely AF
theory for  smaller, $ N_F > 3 $, whereas, in the 
absence of the gauge coupling we had an AF solution 
only for $ N_F > 4 $, where $N_F $ is the number 
of flavours \cite{long}.

 2.{\it Is the Standard Model Higgs Composite?}
 It is a simple exercise to read off the compositeness 
condition from the value of $ \rho_c $. If  $\rho < \rho_c $ then the theory is AF and the higgs elementary. If
on the other hand $ \rho > \rho_c $, the higgs is
composite. It follows, from substituting the known 
top quark yukawa coupling (the other yukawa couplings 
may be neglected as they are much smaller in 
comparison) and the QCD and weak gauge couplings at 
the weak scale, that, even on including the weak 
coupling contribution to  $\rho_c$, we are still
left with a {\em composite higgs}. A full analysis of 
this will be presented elsewhere.
\section { Conclusions.}
\noindent
We have found interesting structure in field theories 
with scalars and gauge fields. Whereas the AF of the 
gauge coupling is gained for large $N_C$ and minimal 
$N_G$, the AF for the yukawa coupling can happen only
when  $ N_G > (N_C/2 + 9/(2 N_C))$, for only then can 
we have a positive $ \rho_c$ (inclusion of the weak 
gauge coupling modifies this condition). Therefore, 
there is a window in which AF for all couplings is 
possible.

To sum up we have found such a window in the above 
theory in which all couplings are AF. This guarantees 
that all particles in the spectrum of the theory
are partonic/ elementary at arbitrarily high energies 
and their wave function renormalization remains 
non-zero. As we go to low energies, however, 
the coloured part of the spectrum is expected to get 
confined into hadrons. Since the scalar/pseudoscalars 
are colour singlets, they are not confined,but
remain part of the low energy spectrum.Thus, we have 
demonstrated that in a renormalizable field theory 
the scalars/pseudoscalars can be elementary particles 
and visible at all scales.
  
Harada etal \cite{harada} have also
looked at the nontriviality of the gauge higgs yukawa 
system and the gauged NJL model. Their emphasis 
is, however, different and not on the issue of
elementarity. Also, they have not considered the effect
of the weak gauge coupling. 

{\bf Acknowledgements}:
Work, of which this forms a part, was started with a 
larger team consisting of Rahul Basu, H.S. Sharatchandra
and Ramesh Anishetty. We thank them for their initial 
contributions. Charan Aulakh, D. Sahdev and 
V. Srinivasan are thanked for animated discussions.
V.S. dedicates this work to his father.

\end{document}